\documentclass[pdftex]{acoust}
\usepackage{bm}
\usepackage{comment}
\usepackage{url}
\usepackage{amsmath,amsfonts}

\def\header{H. OHNAKA {\it{et al.}}: SPEECH ENHANCEMENT WITH KURTOSIS AND DEEP PRIORS} 

\title{Unsupervised speech enhancement with spectral kurtosis and double deep priors}

\author{
Hien Ohnaka$^{1,2}$\thanks{e-mail: onaka.hien.oj5@naist.ac.jp} and 
Ryoichi Miyazaki$^1$\thanks{e-mail: miyazaki@tokuyama.ac.jp}}
\inst{
    $^{1}$ National Institute of Technology, Tokuyama College, Gakuendai, Shunan-shi, Yamaguchi, 745-8585, Japan \\
    $^{2}$ Nara Institute of Science Technology, Takayama-cho, Ikoma-shi, Nara, 630-0101, Japan
}

\abst{
This paper proposes an unsupervised DNN-based speech enhancement approach founded on deep priors (DPs). Here, DP signifies that DNNs are more inclined to produce clean speech signals than noises. Conventional methods based on DP typically involve training on a noisy speech signal using a random noise feature as input, stopping training only a clean speech signal is generated. However, such conventional approaches encounter challenges in determining the optimal stop timing, experience performance degradation due to environmental background noise, and suffer a trade-off between distortion of the clean speech signal and noise reduction performance. To address these challenges, we utilize two DNNs: one to generate a clean speech signal and the other to generate noise. The combined output of these networks closely approximates the noisy speech signal, with a loss term based on spectral kurtosis utilized to separate the noisy speech signal into a clean speech signal and noise.  The key advantage of this method lies in its ability to circumvent trade-offs and early stopping problems, as the signal is decomposed by enough steps. Through evaluation experiments, we demonstrate that the proposed method outperforms conventional methods in the case of white Gaussian and environmental noise while effectively mitigating early stopping problems.
}

\kword{Speech enhancement, Unsupervised learning, Spectral kurtosis, Deep prior, Deep neural network}

\mladdress{National Institute of Technology, Tokuyama College, Gakuendai, Shunan-shi, Yamaguchi, 745-8585, Japan}

\class{7} 

\recommend{1}  

\begin{document}
\maketitle

\section{Introduction} 
The intrusion of noise into speech processing systems is detrimental, as it adversely affects comprehension and degrades the overall quality of the speech signals.
To counteract this, various speech enhancement methods have been developed to remove noise from affected speech signals~\cite{SS,WF,MMSESTSA,SE_DNN,SE_DAE,SE_SEGAN,MixIT,PULSE,NyTT,MetricGANU,DAP_harmo,DAPmus,DAP_wave,DAP_dense,demucs,mlnfree}.
There is a growing consensus in the field that deep neural network (DNN)-based supervised speech enhancement techniques, underpinned by extensive datasets containing both clean and noisy speech signals, can deliver superior performance~\cite{SE_DNN,SE_DAE,SE_SEGAN}.
However, a significant impediment to this approach is the challenge of compiling a comprehensive dataset, particularly due to the prerequisites for recording clean speech signals in anechoic conditions.

Recent scholarly endeavors have been directed towards the development of DNN-based speech enhancement methods, circumventing the necessity for clean speech signals. 
A prominent strategy is the deployment of self-supervised learning frameworks for training speech enhancement DNNs, leveraging extensive datasets of noisy speech signals~\cite{MixIT,PULSE,NyTT,MetricGANU}. 
Notably, noisy-target training~\cite{NyTT} engages in pairing noisy speech signals with additional noises, training the DNN to filter out such disturbances, a skill further applied to cleansing incoming noisy speech signals during inference phases. 
Concurrently, MetricGAN-U~\cite{MetricGANU} represents an innovative approach, integrating a non-intrusive evaluation metric within its loss function, thus streamlining the training mechanism for speech enhancement DNNs. 
Another research vein probes into unsupervised speech enhancement methods using DNNs, eliminating the need for pretraining~\cite{DAP_harmo,DAPmus,DAP_wave,DAP_dense}. 
These approaches are influenced by the deep image prior (DIP)~\cite{DIP} notion in computer vision, harnessing the natural capabilities of untrained DNN frameworks.

\begin{figure}[tb]
\begin{center}
\includegraphics[width=6.5cm]{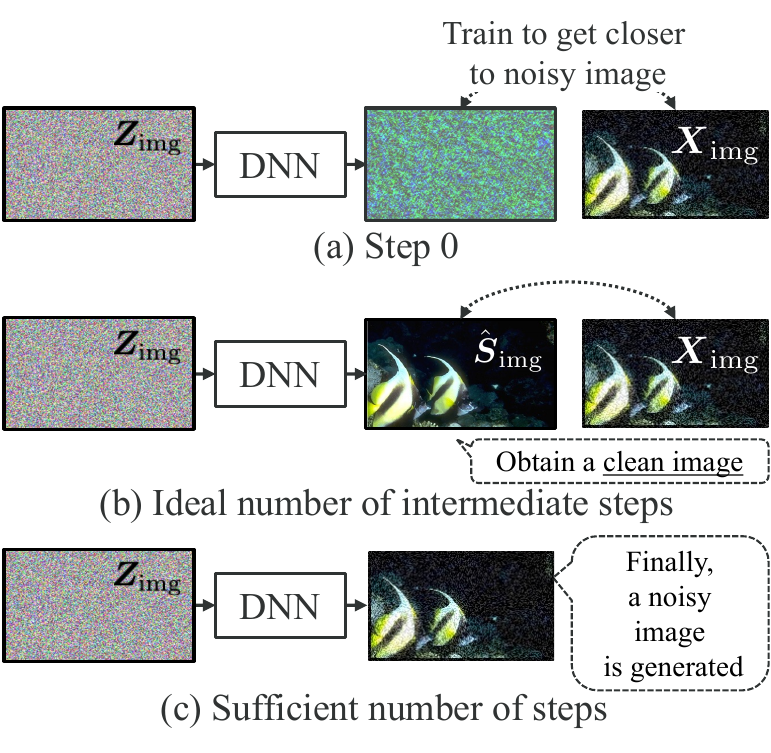}
\caption{Conceptual diagram of image denoising via deep image prior~\cite{DIP}.}
\label{fig:dip_concept}
\end{center}
\end{figure}

DIP is the phenomenon in which an untrained convolutional neural network (CNN) exhibits a predisposition towards generating structured, coherent images as opposed to random noise.
This principle facilitates noise attenuation in the context of training iterations focused on a single noisy image.
In Fig.~\ref{fig:dip_concept}, the strategic cessation of training at an opportune juncture yields a clean image.
Extending this framework to the domain of audio signal processing has validated its utility in speech enhancement~\cite{DAP_dense,DAP_harmo,DAP_wave,DAPmus}. 
Notably, DNNs utilizing harmonic convolution\cite{DAP_harmo} (customized for speech's inherent features) and those integrating dilated convolution and dense connections~\cite{DAP_dense} have succeeded in speech enhancement by training on complex spectrograms. 
Furthermore, Turetzky et al.'s study ~\cite{DAP_wave} delineates the effectiveness of time domain deep prior (DP)-based speech enhancement by utilizing the demucs model~\cite{demucs}, a DNN model dedicated to sound source separation.

Conventional speech enhancement approaches leveraging the DP attributes of untrained DNNs manifest inherent limitations. 
Within this paradigm, the training flow initially produces a clean output followed by a noisy output, thus complicating the identification of an optimal stopping point in non-oracle environments. 
While existing research has substantiated the utility of this technique against white Gaussian noise, its generalizability to diverse acoustic conditions necessitates further empirical validation. 
Our preliminary investigations have revealed pronounced degradation in speech enhancement capabilities within environments exhibiting power gradients across frequency spectral. 
Additionally, there is an inherent trade-off between noise reduction efficiency and speech signal fidelity, which represents a consequential barrier to the advancement of speech enhancement performances.

This study elucidates the challenges in speech enhancement, drawing inspiration from Double-DIP~\cite{double-dip}, an innovative image decomposition approach leveraging double deep priors. 
Utilizing a dichotomous DNN framework, one network is optimized for the clean speech signal generation while the other targets noise production. 
This is augmented by a loss function predicated on spectral kurtosis, enhancing the demarcation between clean speech signals and noise elements. 
The integration of these DNNs facilitates superior speech enhancement, negating the requirement for premature training discontinuation and addressing the balance between effective noise mitigation and speech clarity preservation. 
Comparative analysis reveals that the proposed method substantially surpasses traditional DP-based techniques in reducing a diverse type of noise.

\section{Problem setting}
\subsection{Speech signal formulation}
In the discrete-time domain, the representation of noisy speech signal $\bm{x}=(x[l])_{(l=0)}^{L-1} \in \mathbb{R}^L$ of length $L$ is articulated as a sum of clean speech signal $\bm{s}$ and noise $\bm{n}$, as
\begin{align}
     \bm{x} = \bm{s} + \bm{n},
\end{align}
where $\bm{s}=(s[l])_{(l=0)}^{L-1} \in \mathbb{R}^L$ denotes the clean speech signal and $\bm{n}=(n[l])_{(l=0)}^{L-1} \in \mathbb{R}^L$ represents the noise.
Utilizing the short-time Fourier transform (STFT) with the window function $\bm{w} = (w[l])_{(l=0)}^{W-1} \in \mathbb{R}^W$ of window length $W$, we obtain the real-valued complex spectrogram:
\begin{align}
    \bm{X} &= (\mathrm{Re}[X[k,\tau]], \mathrm{Im}[X[k,\tau]])_{(k=0, \tau=0)}^{K-1,T-1} \in \mathbb{R}^{2 \times K \times T},
\end{align}
\begin{align}
    {X}[k,\tau]&= \sum_{n=0}^{W-1} x_\tau[n] e^{-j 2\pi k n /W}, x_\tau[a] = w[a] x[a+A\tau].
\end{align}
Here, $k$ is the frequency bin index, $\tau$ is the time frame index, and $A$ represent the shift length, respectively.
Hereafter, real-valued complex spectrograms of $\bm{s}$ and $\bm{n}$ will also be denoted by $\bm{S}$ and $\bm{N}$, respectively.

\subsection{DP-based speech enhancement}
Speech enhancement endeavors to extract the noise component $\bm{n}$ from the noisy speech signal $\bm{x}$, aiming to recover the intended clean speech signal $\bm{s}$.
Within the STFT domain, DP-based speech enhancement~\cite{DAP_dense,DAP_harmo} is implemented through the application of the following training equation:
\begin{align}
    \underset{\theta}{\mathrm{min}}\ \mathcal{L} \left(g_{\theta}(\bm{Z}), \bm{X} \right),  \label{eq:dap_optim}
\end{align}
where $\bm{Z}=(Z[i,k,\tau])_{(i=0, k=0, \tau=0)}^{1,K-1,T-1} \in \mathbb{R}^{2 \times K \times T}$ represents the input feature sampled from the normal distribution $\mathcal{N}(0,1)$.
Therefore, $g_{\theta}(\cdot)$ represents the DNN with parameters $\theta$.
This approach leverages a distinctive characteristic wherein $\bm{X}$ is adequately generated after $t_x$ steps, whereas $\bm{S}$ materializes in a reduced number of steps, $t_s$, satisfying  $t_s < t_x$ during training in Eq.~(\ref{eq:dap_optim}).
Terminating the training at $t_s$ facilitates the prediction of a clean speech spectrogram:
\begin{align}
    \hat{\bm{S}} = g_{\theta(t_s)}(\bm{Z}).
\end{align}
Similar methods for achieving speech enhancement in the time-domain~\cite{DAP_wave} and amplitude spectrogram~\cite{DAPmus} have followed a comparable approach.

\section{Motivation} \label{sec:Motivation}
Conventional DP-based speech enhancement methods have primarily been evaluated through experiments in environments containg white Gaussian noise, with only limited scrutiny applied to alternative noise scenarios~\cite{DAP_harmo,DAP_dense}.
Consequently, in our initial investigation, we executed preliminary experiments to determine the efficacy of these methods within environmental noise settings.
In addition, an experiment applying clean speech signals to the target is also conducted (i.e. $\bm{X}$ is replaced by $\bm{S}$ in Eq.~(\ref{eq:dap_optim})). 
Our reason for incorporating clean speech signals in addition to noisy speech signals in this experiment was to assess the extent of distortion each network imparts to the speech signal.

In the preliminary, we utilized 20 1.5-second clips of clean speech signals from the JNAS corpus~\cite{JNAS}.
Two noise types were used: white Gaussian noise and station noise, the latter representing the environmental noise of a busy subway station~\cite{DEMAND}.
The target data comprised 60 samples, including 40 samples of noisy speech mixed to achieve a signal-to-noise ratio (SNR) of 10 dB, and 20 samples of clean speech. 
We utilized the following four comparison methods:
\begin{itemize}
    \item \textbf{U-net}: A normal convolutional layer-based U-net~\cite{U-net}.
    \item \textbf{Harmonic U-net}: A harmonic convolutional layer-based U-net~\cite{DAP_harmo}. The implementation of harmonic convolution is based on harmonic lowering~\cite{HL}.
    \item \textbf{Dense connection and Dilated convolution (DD) U-net}: A U-net incorporating dense connections and dilated convolution~\cite{DAP_dense,DAP_dense_imple}.
    \item \textbf{Deep waveform prior (DWP)}: DP-based speech enhancement utilizing demucs~\cite{DAP_wave,DAP_wave_imple}.
\end{itemize}
For U-net and Harmonic U-net, the U-net consists of five blocks at a depth of two. Each block involves two convolutional layers, followed by instance normalization~\cite{InstanceNorm} and LeakyReLU~\cite{LeakyReLU} activations. The number of channels and down/upsampling for each layer is as follows: 2$\rightarrow$35, 35$\rightarrow$35, average pooling, 35$\rightarrow$70, 70$\rightarrow$70, average pooling, 70$\rightarrow$70, 70$\rightarrow$70, bilinear upsampling, 140$\rightarrow$35, 35$\rightarrow$35, bilinear upsampling, 70$\rightarrow$35, 35$\rightarrow$35.
The final layer consists of a $1\times1$ convolution layer, mapped to the same number of channels as the input.
The kernel size for both methods is $3\times3$, and a mixing process is applied for Harmonic convolution.
The complex spectrogram was derived by applying an STFT with the window length of 512, and shift length of 128.
The Perceptual Evaluation of Speech Quality (PESQ)~\cite{PESQ}, a metric for assessing speech quality, was utilized for evaluation.

\begin{figure}[tb]
\begin{center}
\includegraphics[width=6.5cm]{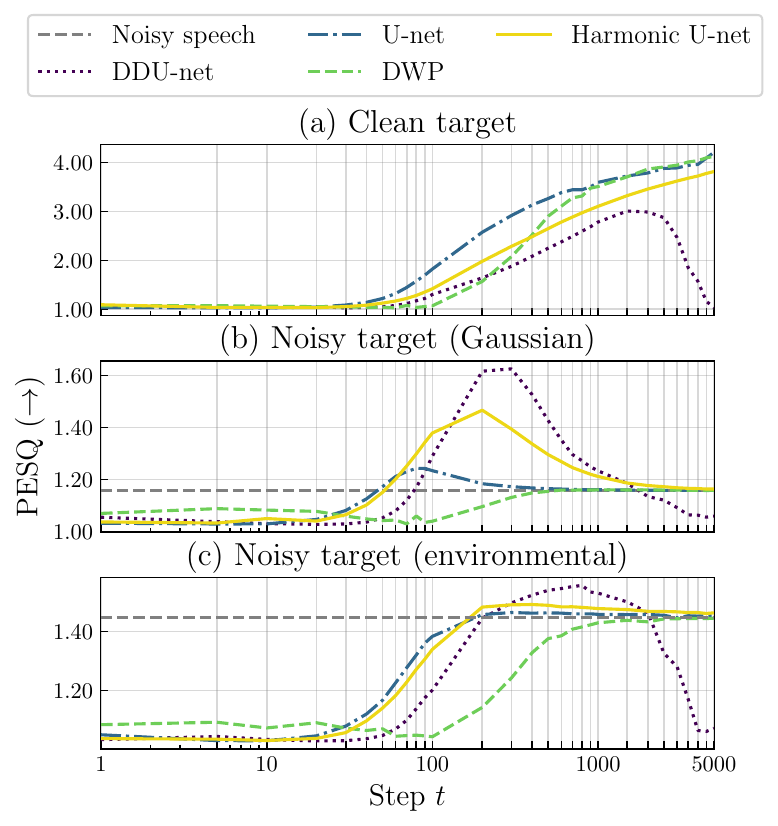}
\caption{A graph of the PESQ~\cite{PESQ} score at each step.}
\label{preexpr_pesq}
\end{center}
\end{figure}

In Fig.~\ref{preexpr_pesq}, the progression of the PESQ scores is graphically represented at each step to assess the early stopping issue and to analyze the signal generation across different network architectures.
As we can see from the analysis of the clean speech targets in Fig.~\ref{preexpr_pesq}(a), analysis of the clean speech targets demonstrates that the U-net and DWP outscored DDU-net and Harmonic U-net.
Conversely in {Fig.~\ref{preexpr_pesq}} (b), when assessing noisy speech targets with white Gaussian noise, the results favored the latter methods.
Notably, DDU-net and Harmonic U-net, both of which are equipped with superior deep priors for speech enhancement, effectively mitigated noise yet induced substantial distortion in clean speech signals.
This distortion presents a considerable impediment, especially under conditions of low ambient noise. 
Moreover, determining the optimal stop timing of processing is difficult without the predictive insights typical of oracle conditions.

\begin{figure}[tb]
\begin{center}
\includegraphics[width=6.5cm]{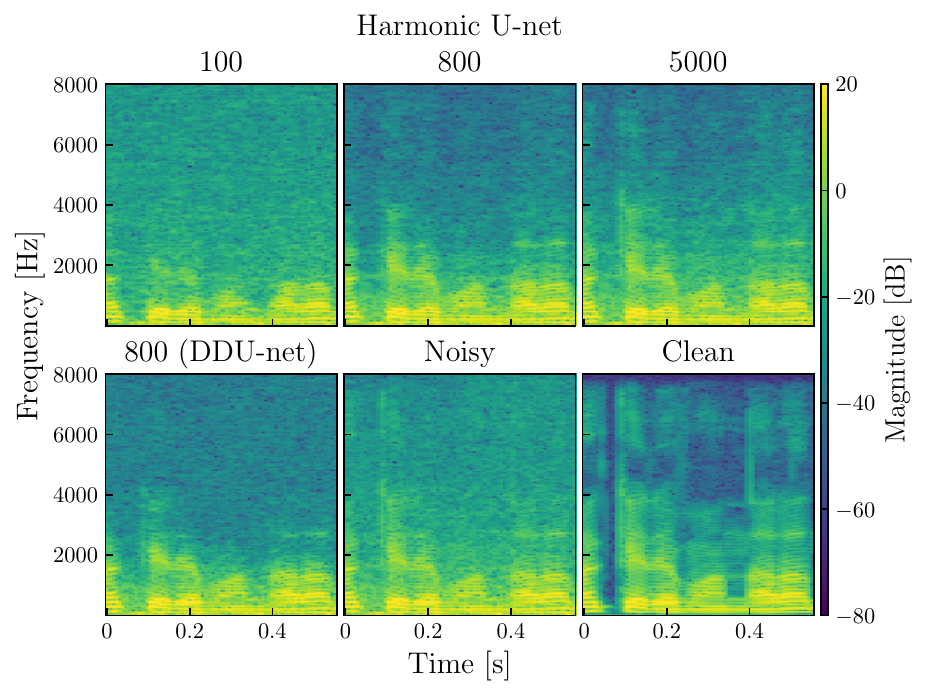}
\caption{Spectrogram of training results for noisy speech with environmental noise using DDU-net.}
\label{preexpr_spec}
\end{center}
\end{figure}
In the case of noisy speech targets subjected to environmental noise, as indicated in {Fig.~\ref{preexpr_pesq}}(c), the resultant PESQ scores are consistently lower than those encountered in white Gaussian noise case. 
{Figure ~\ref{preexpr_spec}} displays sample spectrograms for the environmental noise situation.
At step 100, main noise components below 2000~Hz were not generated, and only a rough form of clean speech signals was captured.
Upon further training, at step 800, the clean speech signal was entirely generated up to the high frequencies, but noise was also generated. 
This outcome stems from the environmental noise is more structural compared to Gaussian noise, which also leads to their preferential generation. 
Consequently, the disparity between $t_s$ and $t_x$ diminishes, resulting in poor performance.

\section{Proposed method}
In the preliminary experiments detailed in Sec.~\ref{sec:Motivation}, we identified deficiencies in established DP-based speech enhancement frameworks, notably, the compromise between noise reduction efficacy and the preservation of clean speech fidelity, alongside inadequate performance in environmental noise scenarios. 
In response, we introduce a new approach that utilizes two designed DNNs: one generated for clean speech generation and the other for noise generation. 
This double-network architecture is trained to generate noisy speech by aggregating their respective outputs, a concept inspired by the Double-DIP approach from computer vision~\cite{double-dip}.
Further, we propose a loss function predicated on spectral kurtosis, which is designed to refine the output of each network and facilitate both noise and clean speech generation. 

\subsection{Overview}
\begin{figure}[t!]
    \begin{center}
    \includegraphics[width=6.5cm]{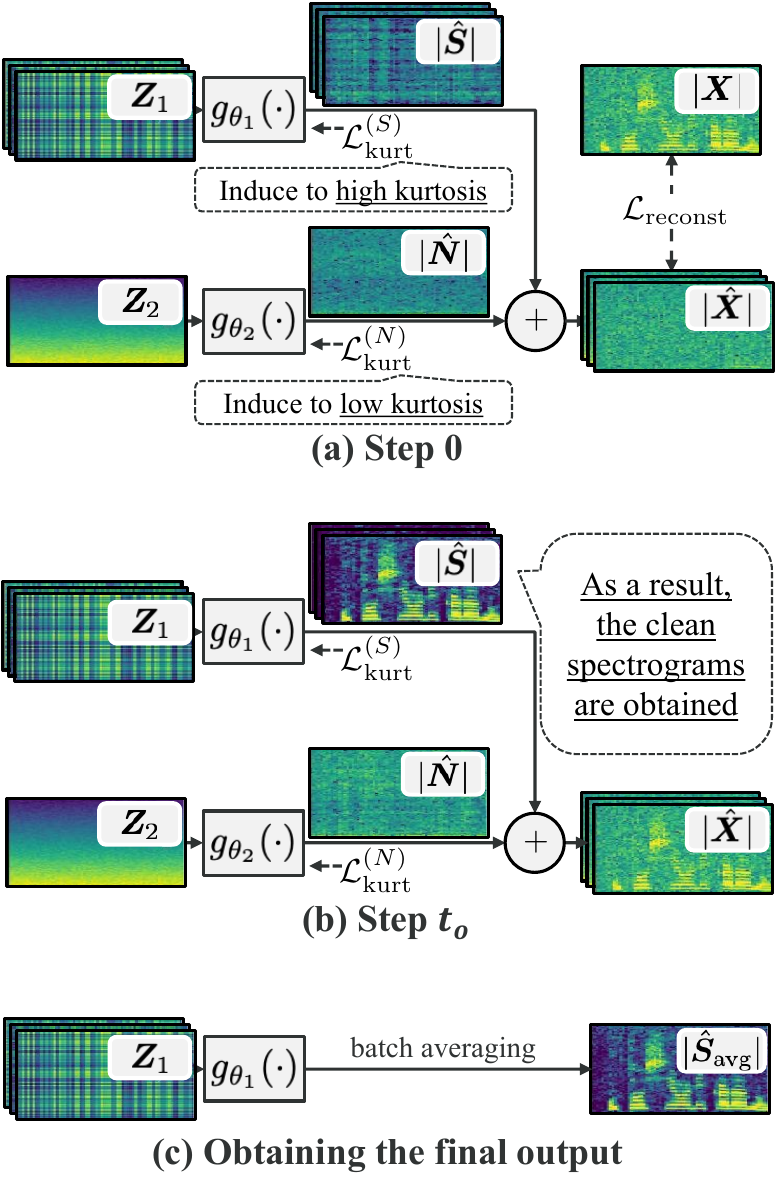}
    \caption{
        Concept of proposed method.
        (a) DNNs are trained so that the sum of $g_{\theta_1}(\bm{Z}_1), g_{\theta_2}(\bm{Z}_2)$ approaches $|\bm{X}|$ and the kurtosis of $|\hat{\bm{S}}|$ is high and $|\hat{\bm{N}}|$ is low.
        (b) After a sufficient number of step $t_o$ iterations, $M$ predicted clean speech signals are obtained.
        (c) The final result is a batch average of predicted clean speech signals.
    }
    \label{fig:prop_concept}
    \end{center}   
\end{figure}

The proposed method adopts amplitude spectrograms as acoustic features, eschewing complex spectrograms or raw waveforms. 
This choice is motivated by the expectation that amplitude spectrograms would be easier to train due to their clearer structure for the clean speech signal and noise~\cite{ipcSpec}.
Additionally, the optimization process is designed to be more straightforward by restricting it to aggregating non-negative features.
The amplitude spectrogram is defined as follows:
\begin{align}
    |\bm{Y}| = (\sqrt{\mathrm{Re}[Y[k,\tau]]^2+\mathrm{Im}[Y[k,\tau]]^2})_{(k=0, \tau=0)}^{K-1,T-1} \in \mathbb{R}^{K \times T}, \label{eq:ampspec}
\end{align}

Initially, we assume that the additivity expressed in Eq. (\ref{eq:approx_add}) holds for the amplitude spectrograms $|\bm{S}|, |\bm{N}|$ as
\begin{align}
    |\bm{X}| \simeq |\bm{S}| + |\bm{N}|. \label{eq:approx_add}
\end{align}
Here, $|\bm{S}|$ and $|\bm{N}|$ are obtained by applying Eq. (\ref{eq:ampspec}) to $\bm{S}$ and $\bm{N}$.
Therefore, we design the noisy spectrogram to be predicted by the sum of two DNNs, as illustrated in Fig.~\ref{fig:prop_concept}.
In this setup, $g_{\theta_1}(\cdot)$ is expected to predict the clean spectrogram, while $g_{\theta_2}(\cdot)$ is expected to predict the noise spectrogram.
Additionally, for $g_{\theta_1}(\cdot)$, the input feature $\bm{Z}_1=(Z_{1}[m,k,\tau])_{(m=0,k=0,\tau=0)}^{M-1,K-1,T-1} \in \mathbb{R}^{M\times K\times T}$ with a batch size of $M$ is utilized for application to batch processing.
For $g_{\theta_2}(\cdot)$, the input feature $\bm{Z}_2=(Z_{2}[k,\tau])_{(k=0,\tau=0)}^{K-1,T-1} \in \mathbb{R}^{K\times T}$ is used.
The designs of the DNNs and input features are detailed in Sec.~\ref{sec:design_dp}.
In consideration of these specifications, Eq. (\ref{eq:approx_add}) is simulated by two DNNs as follows:
\begin{align}
    |\hat{\bm{X}}| &= (|\hat{{X}}[m,k,\tau]|)_{(m=0,k=0,\tau=0)}^{M-1,K-1,T-1}, \\
    |\hat{{X}}[m,k,\tau]| &= |\hat{{S}}[m,k,\tau]| + |\hat{{N}}[k,\tau]|, \\ 
    |\hat{\bm{S}}|&=g_{\theta_1}(\bm{Z}_{1}), |\hat{\bm{N}}|=g_{\theta_2}(\bm{Z}_2).
\end{align}
Training is conducted using Eq. (\ref{eq:prop_optim}).
\begin{align}
    \underset{\theta_1, \theta_2}{\mathrm{min}}\    \mathcal{L}_\mathrm{kurt}^{(S)}(|\hat{\bm{S}}|,|{\bm{X}}|) + \mathcal{L}_\mathrm{kurt}^{(N)}&(|\hat{\bm{N}}|,|{\bm{X}}|) \nonumber \\
    & + \mathcal{L}_\mathrm{reconst}(|\hat{\bm{X}}|,|\bm{X}|),\label{eq:prop_optim}
\end{align}
\begin{align}
   \mathcal{L}&_\mathrm{reconst}(|\hat{\bm{X}}|,|\bm{X}|)\notag\\
   &=\frac{1}{MKT}\sum_{m=0}^{M-1}\sum_{k=0}^{K-1}\sum_{\tau=0}^{T-1}\left||\hat{{X}}[m,k,\tau]|-|{X}[k,\tau]|\right|. 
\end{align}
Here, $\mathcal{L}_\mathrm{kurt}^{(S)}$ and $\mathcal{L}_\mathrm{kurt}^{(N)}$ denote loss terms designed to encourage the decomposition of the clean and noise spectrograms, with further details explained in Sec.~\ref{sec:kurtloss}.
$\mathcal{L}_\mathrm{reconst}$ represents the reconstruction error between $|\hat{\bm{X}}|$ and $|\bm{X}|$, using the mean absolute error.
With these innovations, after a sufficient number of steps $t_o$, the estimated clean spectrograms and noise spectrogram are obtained, as depicted in {Fig. ~\ref{fig:prop_concept}}(b).
A key advantage of our method is the elimination of the need to determine $t_s$ as in conventional methods. 
This is because our method only requires training enough steps $t_o$ to ensure that $g_{\theta_1}(\cdot)$ generate clean spectrograms and $g_{\theta_2}(\cdot)$ is output as a noise spectrogram.
The final output is the batch-averaged clean spectrogram $|\hat{\bm{S}}_\mathrm{avg}| = (\frac{1}{M}\sum_{m=0}^{M-1} |\hat{{S}}[m,k,\tau]|)_{(k=0,\tau=0)}^{K-1,T-1}$ ({Fig.~\ref{fig:prop_concept}}(c)).

\subsection{Design of deep priors} \label{sec:design_dp}
We delineate two fundamental elements in the architectural design of deep priors. 
Let $t_{g1}$ denotes the number of steps for $g_{\theta_1}(\cdot)$ to generate sufficiently accurate target spectrograms, and let $t_{g2}$ denote the number of steps $g_{\theta_2}(\cdot)$ to generate certain spectrograms.
The following relationships are crucial:
\begin{itemize}
    \item For training a clean spectrogram, $t_{g1} < t_{g2}$.
    \item For training a noise spectrogram, $t_{g2} < t_{g1}$.
\end{itemize}
These relationships are paramount because, under such conditions, the generation of clean spectrograms is expected to be primarily directed towards $g_{\theta_1}(\cdot)$, while the generation of noise spectrograms is focused on $g_{\theta_2}(\cdot)$.

To obtain these relationships, we carefully design the input features $\bm{Z}$ and the output layer. 
Previous studies (e.g., video processing as observed in Double-DIP~\cite{double-dip}, and in source separation methods inspired by Double-DIP~\cite{DAPsepa}) has demonstrated that maintaining consistent values of input random features in the time direction can enhance the consistency of output in the time direction. 
Inspired by these precedents, we define $\bm{Z}_{1,m}$ following the parameters in Eq.~(\ref{eq:z1}) to generate a clean spectrogram characterized by uniformity across both time and frequency axes.
\begin{align}
    Z_{1}[m,k,\tau] &= \frac{1}{2}(u_{m,k} + u_{m,\tau}). \label{eq:z1}
\end{align}
Here, $u_{m,k}$ and $u_{m,\tau}$ represent input random numbers sampled from $U(0,0.1)$, respectively.
$u_{m,k}$ exhibits variability across frequency bins while maintaining consistent values across time frames, whereas $u_{m,\tau}$ shows the variability across time frames while holding values constant across frequency bins.
As a result, this feature visually appears as a line in both the time and frequency dimensions, as illustrated by $\bm{Z}_{1}$ in {Fig.~\ref{fig:prop_concept}}.
For $\bm{Z}_2$, we utilize a mesh grid whose values gradually decrease from the low-frequency side to the high-frequency side, as illustrated by $\bm{Z}_2$ in {Fig.~\ref{fig:prop_concept}}.
Ulyanov demonstrated that a mesh grid serves as a prior distribution that encourages smooth output~\cite{DIP}. 
Consequently, it is anticipated to function as a smooth noise signal characteristic. In our proposed method, 
$\bm{Z}_2$ is defined by Eq.~(\ref{eq:z2}), which comprises the meshgrid term plus a small perturbation $u_{k,\tau}$.
\begin{align}
    Z_2[k,\tau] &= 0.09\frac{K-k}{K} + 0.01u_{k,\tau}. \label{eq:z2}
\end{align}
Here, $u_{k,\tau}$ is a random number sampled from $U(0,0.1)$.
We designed both $\bm{Z}_{1}$ and $\bm{Z}_{2}$ to have a range of possible values between $0$ and $0.1$.

\begin{figure}[t!]
\begin{center}
\includegraphics[width=6.5cm]{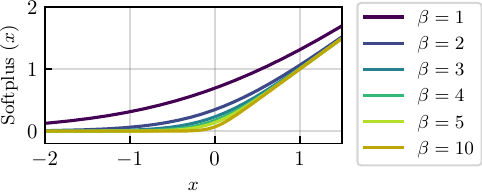}
\caption{
    Graph of softplus function.
}
\label{fig:softplus}
\end{center}   
\end{figure}
Additionally, each output layer of the DNNs leverages a softplus function tailored to match the sparsity characteristics of the respective signals.
The clean spectrogram $|\bm{S}|$ typically represents sparsity, characterized by a minority of high-amplitude components indicative of speech and a majority of minimal amplitude components reflecting silence.
Conversely, $|\bm{N}|$ is a non-sparse signal.
As depicted in {Fig.~\ref{fig:softplus}}, the graph shape of the softplus function approaches that of ReLU with increasing values of the parameter $\beta$.
In our proposed method, we assign a softplus function with different parameters is assigned to each DNN, considering the nature of each signal. 
Specifically, $g_{\theta_1}(\cdot)$ applies a high-beta softplus function to effectively capture the sparsity of $|\bm{S}|$, whereas $g_{\theta_2}(\cdot)$ utilizes a low-beta softplus function to better represent the characteristics of $|\bm{N}|$.

\begin{figure}[t!]
\begin{center}
\includegraphics[width=6.5cm]{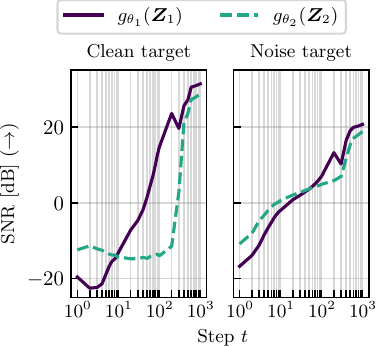}
\caption{
    SNR progression of two DNNs $g_{\theta_1}(\cdot),g_{\theta_2}(\cdot)$ for clean speech target and noisy speech target with white Gaussian noise, respectively. Note that the batch size $M$ of $\bm{Z}_1$ is one.
}
\label{fig:optim_deepprior}
\end{center}   
\end{figure}
To assess the impacts of the input features and output layers described above, we present the learning outcomes for clean and noise spectrograms by combining different $\beta$ softplus functions and $\bm{Z}_1,\bm{Z}_2$ in {Fig.~\ref{fig:optim_deepprior}}.
In the learning process for the clean spectrogram (depicted in Fig. \ref{fig:optim_deepprior}(a)), there is a noticeable trend of score escalation after a certain number of steps, exhibiting a faster generation rate.
Additionally, fewer steps are required for $g_\theta(\bm{Z}_1)$ than for $g_\theta(\bm{Z}_2)$.
This result indicates that $t_{g1} < t_{g2}$ holds during clean spectrogram generation.
In contrast, the optimization for the noise spectrogram (shown in {Fig.~\ref{fig:optim_deepprior}}(b)) tends to differ from the clean spectrogram.
Here, the scores increase gradually for both DNNs.
Notably, $g_\theta(\bm{Z}_2)$ achieves a higher score with fewer steps.
This suggests that $t_{g2} < t_{g1}$ in noise spectrogram generation.

\subsection{Loss term based on spectral kurtosis} \label{sec:kurtloss}
Kurtosis is a statistical measure that reflects the sharpness of a distribution.
In acoustic signal processing, kurtosis has been utilized in various frameworks, such as voice activity detection~\cite{vad_kurt} and source separation~\cite{ica_kurt,ica_moment}, as it is useful information for classifying noise and speech (or music).

We focus on computing the kurtosis within each segmented region of the time-frequency domain.
The kurtosis calculation method by assuming a gamma distribution in the power spectrogram of speech signals~\cite{SpecKurt} is extended to the segmented region.
Initially, we define the expected value $\mathcal{E}$ in the segmented region, where the signal is partitioned for each time-frequency direction.
\begin{align}
    &\mathcal{E}_{r_\tau}^{r_k}\{|\bm{Y}|^2\} \notag\\
    &\hspace{1.5mm}= \left(\frac{1}{r_k r_\tau}\sum_{k_r=0}^{r_k-1}\sum_{\tau_r=0}^{r_\tau-1}{|{Y}[r_k \tilde{k} + k_r, r_\tau \tilde{\tau} + \tau_r]}|^2\right)_{\tilde{k}=0,\tilde{\tau}=0}^{\tilde{K}-1,\tilde{T}-1}. \label{eq:conv}
\end{align}
Here, $r_k$ and $r_\tau$ represent the number of elements in one partition in the frequency and time directions, and $\tilde{K}$ and $\tilde{T}$ are the number of partitions in the frequency and time directions, respectively.
The kurtosis $\bm{\mathcal{K}}_{Y(r_k,r_\tau)}$ in each region divided by $r_k$ and $r_\tau$ is calculated according to Eq.~(\ref{eq:spec_kurt2}).
\begin{align}
    \bm{\mathcal{K}}_{Y(r_k,r_\tau)} &= (\mathcal{K}_{Y(r_k,r_\tau)}[\tilde{k},\tilde{\tau}])_{\tilde{k}=0,\tilde{\tau}=0}^{\tilde{K}-1,\tilde{T}-1}, \label{eq:spec_kurt2} \\
    \mathcal{K}_{Y(r_k,r_\tau)}&[\tilde{k},\tilde{\tau}] = \frac{
        (\hat{\eta}_Y[\tilde{k},\tilde{\tau}]+2)
        (\hat{\eta}_Y[\tilde{k},\tilde{\tau}]+3)
    }{
        \hat{\eta}_Y[\tilde{k},\tilde{\tau}]
        (\hat{\eta}_Y[\tilde{k},\tilde{\tau}]+1)
    }, \\
    \hat{\bm{\eta}}_Y &= \frac{3-\hat{\bm{\gamma}}_Y +\sqrt{(\hat{\bm{\gamma}}_Y-3)^2+24\hat{\bm{\gamma}}_Y}}{12\hat{\bm{\gamma}}_Y}, \\
    \hat{\bm{\gamma}}_Y &= \log(\mathcal{E}_{r_\tau}^{r_k}\{|\bm{Y}|^2\}) - \mathcal{E}_{r_\tau}^{r_k}\{\log(|\bm{Y}|^2)\}.
\end{align}
Here, $\hat{\bm{\eta}}_Y$ are the estimated shape parameters.

\begin{figure}[t!]
    \begin{center}
    \includegraphics[width=6.5cm]{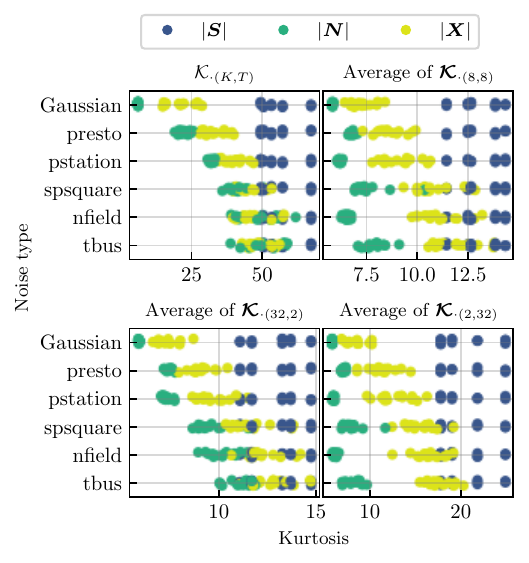}
    \caption{
    Scatter plots of the kurtosis calculated for the entire spectrogram $\mathcal{K}_{\cdot(K,T)}$ and the average kurtosis in the split region $\bm{\mathcal{K}}_{\cdot(8,8)}$, $\bm{\mathcal{K}}_{\cdot(32,2)}$, and $\bm{\mathcal{K}}_{\cdot(2,32)}$.
    }
    \label{fig:scatter_kurt}
    \end{center}    
\end{figure}
A comparison of the kurtosis computed from the spectrograms of various clean, noise, and noisy spectrograms and the kurtosis in the segmented region is illustrated in {Fig.~\ref{fig:scatter_kurt}}. 
Here, the colored points indicate the kurtosis or the average of the kurtosis in the split region for each speech sample.
Note that we set the number of elements $(r_k,r_\tau)$ in one partition to three pairs: $(8,8)$, $(32,2)$, and $(2,32)$. 
The results reveal several key insights. 
First, it is evident that the kurtosis computed for the entire signal exhibits variation depending on the type of noise, akin to previous findings \cite{SpecKurt}. 
This variability arises from dynamic power fluctuations in the frequency and time directions. 
In contrast, the kurtosis in the segmented domain alleviates these dynamic power variations. 
Consequently, the noise signal tends to exhibit a lower value, while the clean signal tends to manifest a higher value, irrespective of the type of noise. 
However, this tendency is very weak for $(32,2)$, which has a large region in the frequency direction. 
This is presumably because the power fluctuation in each frequency band contributes significantly to the kurtosis increase. 
This observation suggests that kurtosis in the segmented domain serves as a valuable indicator for generating noisy spectrograms separately for clean and noise spectrograms.

The segmental spectral kurtosis is directly integrated into the loss term during actual training.
This approach is expected to be similarly effective after successful attempts to incorporate spectral kurtosis into the loss function to regulate the kurtosis of the output signal \cite{DNNkurt}.
The kurtosis-inducing loss term $\mathcal{L}_\mathrm{kurt}^{(S)}$ for $|\hat{\bm{S}}|$ is defined as follows:
\begin{align}
    \mathcal{L}_\mathrm{kurt}^{(S)} &= \mathcal{L}_1^{(S)} + \mathcal{L}_2^{(S)}. \label{eq:kurts_loss}
\end{align}
where $\mathcal{L}_1^{(S)}$ promotes the increase in the kurtosis of each batch output, driving $|\hat{\bm{S}}|$ towards $|{\bm{S}}|$.
$\mathcal{L}_2^{(S)}$ is computed on the batch-averaged signal $|\hat{\bm{S}}_\mathrm{avg}|$ to enhance the speech component and reduce residual noise.
\begin{align}
    \mathcal{L}_1^{(S)} &=
    \frac{-\alpha_1}{\tilde{K}\tilde{T}M}\sum_{\tilde{k}=0}^{\tilde{K}-1} \sum_{\tilde{\tau}=0}^{\tilde{T}-1} \sum_{m=0}^{M-1} 
    \left(
        \frac{ 
            \mathcal{K}_{\hat{S}(r_k,r_\tau)}[m,\tilde{k},\tilde{\tau}]
        }{
            \tilde{\mathcal{K}}_{X(r_k,r_\tau)}[\tilde{k},\tilde{\tau}]
        }
    \right)^2,  \label{eq:kurts1_loss} \\
    \mathcal{L}_2^{(S)} &= 
    \frac{\alpha_2}{\tilde{T}} \sum_{\tilde{\tau}=0}^{\tilde{T}-1}
    \left(
        \frac{
            \mathcal{K}_{\hat{{S}}_\mathrm{avg}(K,r_\tau)}[\tilde{\tau}]
        }{
            \mathcal{K}_{X(K,r_\tau)}[\tilde{\tau}]
        }
    \right)^2 \notag \\
    &\quad\quad -\frac{\alpha_3}{\tilde{K}} \sum_{\tilde{k}=0}^{\tilde{K}-1} 
    \left(
        \frac{
            \mathcal{K}_{\hat{{S}}_\mathrm{avg}(r_k,T)}[\tilde{k}]
        }{
            \tilde{\mathcal{K}}_{X(r_k,T)}[\tilde{k}]
        }
    \right)^2. \label{eq:kurts2_loss}
\end{align}
Here, $\alpha_1,\alpha_2,\alpha_3$ are weight parameters, $\mathcal{K}_{\hat{S}(r_k,r_\tau)}[m,\tilde{k},\tilde{\tau}]$ is the segmented region kurtosis in a certain batch $m$ in $|\hat{\bm{S}}|$, and $\tilde{\bm{\mathcal{K}}}_{X(r_k,r_\tau)}$ denotes the kurtosis in the segmented region, inverted between the largest and smallest values shown in the Eq.~(\ref{eq:inv_kurt}).
\begin{align}
    &\tilde{\bm{\mathcal{K}}}_{X(r_k,r_\tau)} = 
    \max \left( \bm{\mathcal{K}}_{X(r_k,r_\tau)} \right) 
    - \bm{\mathcal{K}}_{X(r_k,r_\tau)}\notag\\
     &\qquad\qquad\qquad\qquad+ \min \left( \bm{\mathcal{K}}_{X(r_k,r_\tau)} \right) . \label{eq:inv_kurt} 
\end{align}
The (inverse) kurtosis of the noisy speech spectrogram in the denominator acts as a weighting factor, giving greater weight to the low (high) kurtosis parts of the original noisy speech spectrogram.

\begin{figure}[t!]
    \begin{center}
    \includegraphics[width=6.5cm]{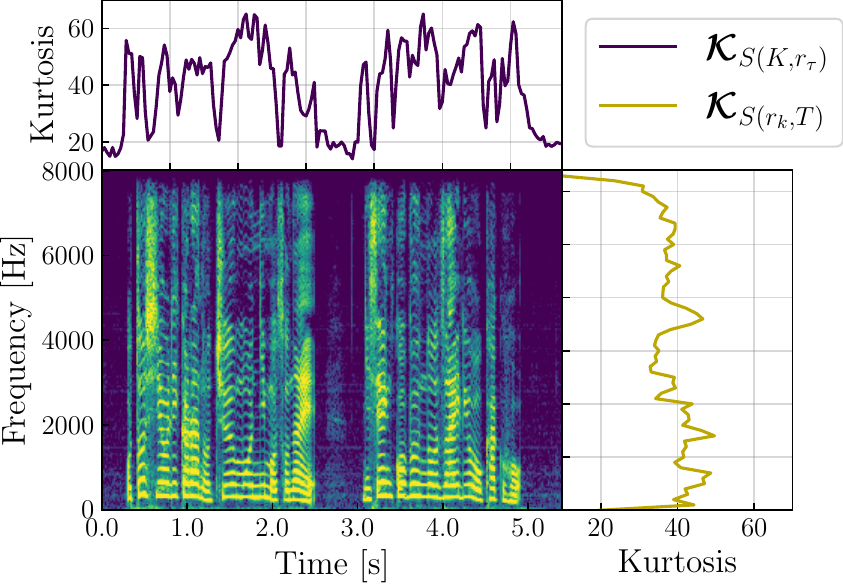}
    \caption{
        Clean spectrogram $|\bm{S}|$, its kurtosis in the short time domain $\bm{\mathcal{K}}_{S(K,r_\tau)}$ and its kurtosis in the subbands $\bm{\mathcal{K}}_{S(r_k,T)}$. Here, $r_k$ and $r_\tau$, was set 4 and 4, respectively.
    }
    \label{fig:clean_kurt}
    \end{center}
\end{figure}

We next describe the design intent of $\mathcal{L}_2^{(S)}$. 
{Figure~\ref{fig:clean_kurt}} shows that $\bm{\mathcal{K}}_{S(K,r_\tau)}$ exhibits low kurtosis during unvoiced segments and high kurtosis during voiced segments. 
The first term of $\mathcal{L}_2^{(S)}$ aims to reduce $\bm{\mathcal{K}}_{\hat{S}_\mathrm{avg}(K,r_\tau)}$. 
This reduction particularly affects unvoiced segments where low kurtosis is desired to reduce the generation of artifact noise.
In contrast, $\bm{\mathcal{K}}_{S(K,r_\tau)}$ in the voice interval is high, and intuitively, the kurtosis reduction seems to have an adverse effect.
However, in fact, the first term of $\mathcal{L}_2^{(S)}$ mitigates the adverse effects of excessive kurtosis increase due to $\mathcal{L}_1^{(S)}$ and reduces speech distortion in the voice section.
The second term of $\mathcal{L}_2^{(S)}$ is designed to retain the high kurtosis exhibited by $\bm{\mathcal{K}}_{{S}(r_k,T)}$ across all frequency bands.
This retention of high kurtosis ensures that the characteristics of $|\bm{S}|$ are effectively preserved in the generated clean spectrogram.

\begin{figure*}[t!]
    \begin{center}
    \includegraphics[width=12.5cm]{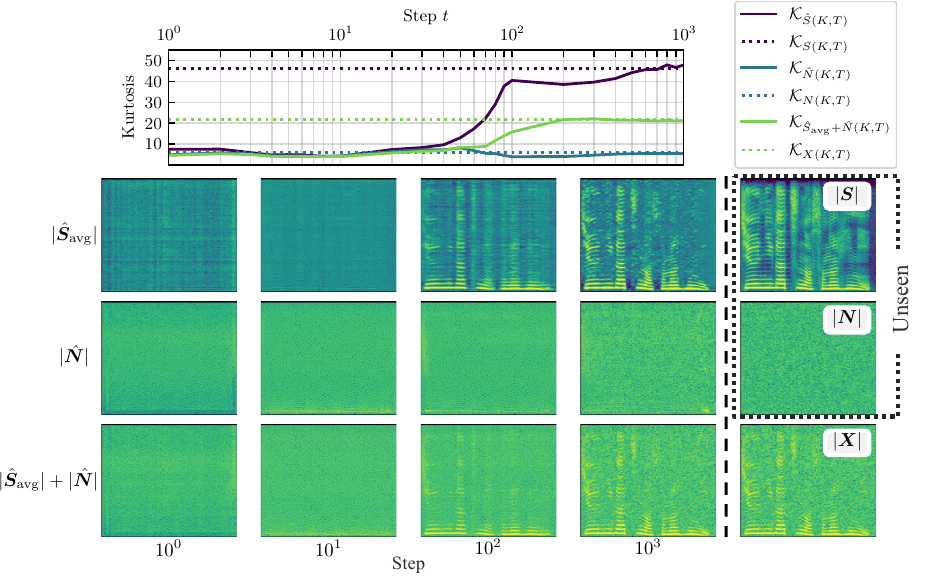}
    \caption{
       Kurtosis transition at learning and spectrograms at 1, 10, 100, and 1000 steps.
       $\mathcal{K}_{\cdot(K,T)}$ is the kurtosis calculated for the entire spectrogram.
    }
    \label{fig:kurtstep_sample}
    \end{center}
\end{figure*}
The loss term $\mathcal{L}_\mathrm{kurt}^{(N)}$, which aims to decrease the kurtosis of $|\hat{\bm{N}}|$, is defined by
\begin{align}
    \mathcal{L}_\mathrm{kurt}^{(N)} &=
    \frac{\alpha_4}{\tilde{K}\tilde{T}}\sum_{\tilde{k}=0}^{\tilde{K}-1} \sum_{\tilde{\tau}=0}^{\tilde{T}-1}
    \left(
        \frac{ 
            \mathcal{K}_{\hat{N}(r_k,r_\tau)}[\tilde{k},\tilde{\tau}]
        }{
            \tilde{\mathcal{K}}_{{X}(r_k,r_\tau)}[\tilde{k},\tilde{\tau}]
        }
    \right)^2, \label{eq:kurtn_loss}
\end{align}
where $\alpha_4$ represents the weight parameter.
$\mathcal{L}_\mathrm{kurt}^{(N)}$ serves to decrease the kurtosis of $|\hat{\bm{N}}|$ and encourages it to approximate $|{\bm{N}}|$.

{Figure~\ref{fig:kurtstep_sample}} illustrates the training outcomes on a noisy spectrogram with white Gaussian noise.
Firstly, the graph shows that $|\hat{\bm{S}}_\mathrm{avg}|$ reaches high values while $|\hat{\bm{N}}|$ approaches the lower values throughout the training process.
Moreover, $|\hat{\bm{X}}|$, which represents the sum of these outputs, progressively aligns with the kurtosis of the noisy input, aligning with our intended objective.
Secondly, the spectrograms reveal that $|\hat{\bm{S}}_\mathrm{avg}|$ converges towards $|{\bm{S}}|$ and $|\hat{\bm{N}}|$ converges towards $|{\bm{N}}|$.
These results confirm that the proposed method can separately generate clean and noise spectrograms from a noisy spectrogram.

\section{Experimental evaluation}
This section describes the results of experiments to evaluate the performance of the proposed method.
In Sec.~\ref{sec:ex_cond}, the experimental conditions are described.
In Sec.~\ref{sec:ex_ovlresult}, results are compared with the competitive methods mentioned in Sec.~\ref{sec:Motivation} to confirm the effectiveness of the proposed method, especially its superior performance against environmental noise.
In Sec.~\ref{sec:ex_earlystop}, we analyze the white Gaussian noise results in more detail and show that the proposed method eliminates the early stopping problem.
Finally, Sec.~\ref{sec:ex_ablation} presents the results of an ablation study that confirms the effectiveness of each component of the proposed method.
\subsection{Experimental conditions} \label{sec:ex_cond}
\noindent\textbf{Test dataset:}
We utilized 100 2-second speech clips from the JNAS corpus~\cite{JNAS} as clean speech signals. 
Additionally, five environmental noises~(presto, pstation, spsquare, nfield, and tbus)~\cite{DEMAND} along with white Gaussian noise were used as noise sources. 
In total, 600 noisy speech clips were evaluated with SNR set to 5, 10, or 15~dB.
\\\textbf{Proposed method: }
DNNs resembling the convolution-based U-net described in Sec.~\ref{sec:Motivation} were utilized for both $g_{\theta_1}(\cdot)$ and $g_{\theta_2}(\cdot)$.
We utilized the Adam optimization algorithm with a learning rate of 0.001.
The weights of the loss functions in the proposed method were set to $(\alpha_1,\alpha_2,\alpha_3,\alpha_4)=(0.00001,0.001,0.00001,2.0)$.
The number of divisions for kurtosis calculation in each loss term in Eqs.~(\ref{eq:kurts1_loss}, \ref{eq:kurtn_loss}), $r_k$ and $r_\tau$, was set to $2$ and $32$, respectively.
In addition, in Eq.~(\ref{eq:kurts2_loss}), $r_k$ and $r_\tau$, was set to $16$ and $16$, respectively.
Noisy phase spectrograms were utilized to obtain waveforms in inverse STFT.
We conducted a comparative analysis of the proposed method against four DP-based speech enhancement methods (as described in Sec.~\ref{sec:Motivation}). 
We used STFT with the window length of 512, and shift length of 128 for both the proposed and comparison methods.
\\\textbf{Objective metrics:}
Three metrics were utilized for evaluation:
\begin{itemize}
    \item Perceptual Evaluation of Speech Quality (PESQ)~\cite{PESQ}, which measures speech quality.
    \item Scale-Invariant Signal-to-Distortion Ratio (SI-SDR)~\cite{sisdr}, which evaluates speech distortion after speech enhancement.
    \item Extended Short-Time Objective Intelligibility (ESTOI)~\cite{estoi}, which represents the intelligibility of speech.
\end{itemize}

\subsection{Evaluation results} \label{sec:ex_ovlresult}
\begin{table*}[tb]
\centering
\label{tb:ovl_score}
\caption{Results on the overall scores}
\scalebox{0.7}{
\tabcolsep = 1pt
\begin{tabular}{l|cccccc|cccccc|cccccc}
& \multicolumn{6}{c|}{SI-SDR} & \multicolumn{6}{c|}{PESQ} & \multicolumn{6}{c}{ESTOI}  \\
& Noisy & U-net & Harmo. & DDU-net & DWP & Prop. & Noisy & U-net & Harmo. & DDU-net & DWP & Prop. & Noisy & U-net & Harmo. & DDU-net & DWP & Prop. \\ \hline
Gaussian & $9.92$ & $11.83$ & $14.41$ & $14.96$ & $10.10$ & $\mathbf{15.10}$ & $1.18$ & $1.26$ & $1.44$ & $1.52$ & $1.18$ & $\mathbf{1.81}$ & $0.758$ & $0.762$ & $0.771$ & $0.753$ & $0.753$ & $\mathbf{0.818}$ \\
presto & $9.91$ & $10.16$ & $10.28$ & $10.54$ & $9.95$ & $\mathbf{12.15}$ & $1.35$ & $1.40$ & $1.44$ & $1.52$ & $1.35$ & $\mathbf{1.62}$ & $0.741$ & $0.737$ & $0.720$ & $0.697$ & $0.740$ & $\mathbf{0.784}$ \\
pstation & $9.92$ & $10.22$ & $10.10$ & $10.07$ & $9.96$ & $\mathbf{13.98}$ & $1.50$ & $1.51$ & $1.54$ & $1.60$ & $1.49$ & $\mathbf{1.95}$ & $0.821$ & $0.815$ & $0.802$ & $0.788$ & $0.819$ & $\mathbf{0.866}$ \\
spsquare & $9.92$ & $10.24$ & $10.05$ & $9.87$ & $9.95$ & $\mathbf{14.84}$ & $1.86$ & $1.81$ & $1.84$ & $1.87$ & $1.82$ & $\mathbf{2.10}$ & $0.895$ & $0.884$ & $0.883$ & $0.866$ & $0.892$ & $\mathbf{0.901}$ \\
nfield & $9.91$ & $10.33$ & $10.05$ & $9.85$ & $9.96$ & $\mathbf{16.76}$ & $2.18$ & $2.04$ & $2.08$ & $2.03$ & $2.08$ & $\mathbf{2.48}$ & $\mathbf{0.937}$ & $0.923$ & $0.923$ & $0.889$ & $0.933$ & $0.932$ \\
tbus & $9.91$ & $10.54$ & $10.17$ & $9.87$ & $9.97$ & $\mathbf{16.14}$ & $\mathbf{2.80}$ & $2.40$ & $2.45$ & $2.40$ & $2.54$ & $2.63$ & $\mathbf{0.961}$ & $0.943$ & $0.934$ & $0.909$ & $0.955$ & $0.931$ \\
\end{tabular}}
\end{table*}

We first discuss the overall results of the proposed and comparative methods.
{Table~1} lists the scores for each type of noise across all methods. 
Note here that {\bf each score was computed for the output with the highest SI-SDR score over 2000 training iterations.} 
Upon inspection of {Table~1}, it becomes evident that the proposed method outperforms all other methods across all conditions, except for the ESTOI scores for nfield and tbus.
These results demonstrate that the proposed method effectively improves the speech enhancement performance. 
Moreover, the proposed method works well even under environmental noise conditions, where the conventional methods typically yield lower scores. 
This robustness is attributed to the innovative use of double DNNs to resolve the typical trade-offs inherent in DP-based speech enhancement practices.

\begin{figure}[t!]
\begin{center}
\includegraphics[width=6.5cm]{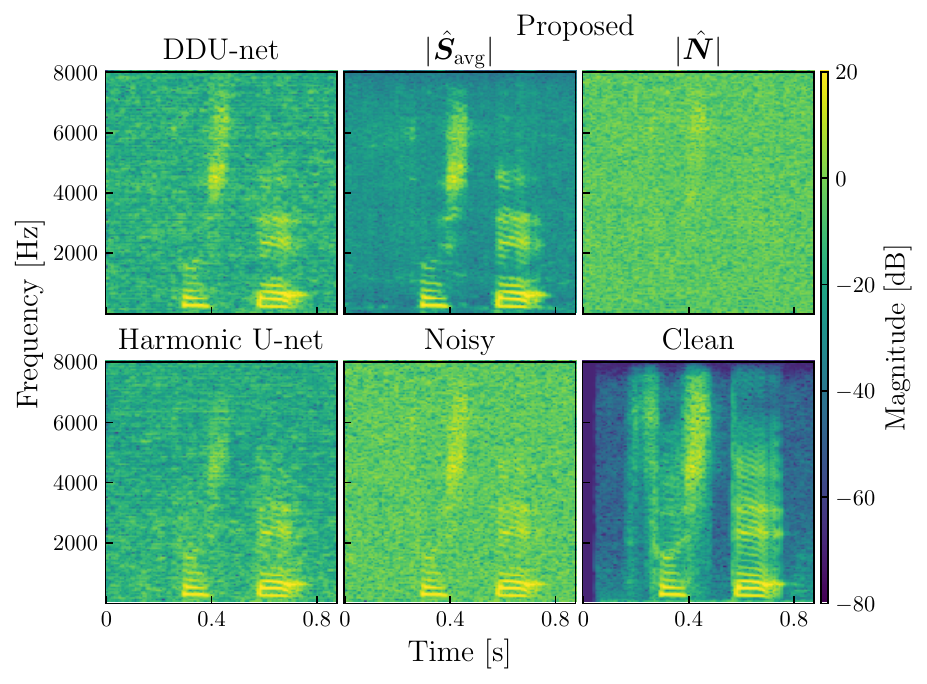}
\caption{
     Processed samples for noisy speech with white Gaussian noise.
}
\label{fig:gauss_spec}
\end{center}   
\end{figure}
\begin{figure}[t!]
\begin{center}
\includegraphics[width=6.5cm]{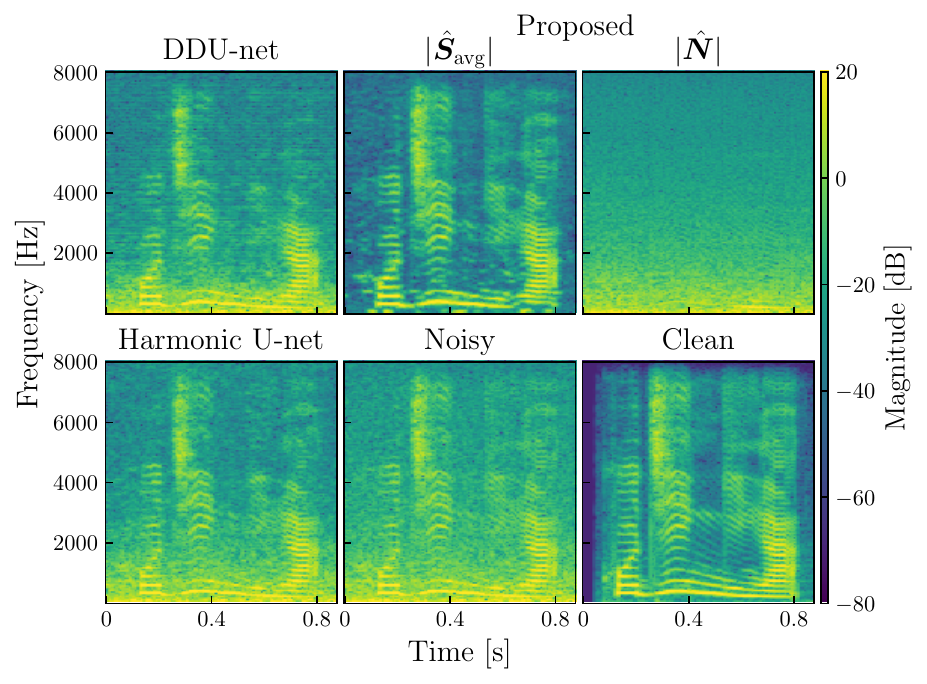}
\caption{
     Processed samples for noisy speech with pstation noise.
}
\label{fig:pstation_spec}
\end{center}   
\end{figure}
These results are also supported in the spectrograms. 
In {Fig.~\ref{fig:gauss_spec}}, the outcomes with white Gaussian noise indicate that the proposed method generates a clean speech signal to a similar or superior extent compared to the conventional methods while exhibiting reduced noise levels. 
Moreover, the results in {Fig.~\ref{fig:pstation_spec}}, which show the outcomes with pstation noise, indicate that the proposed method effectively generates clean speech components in the high-frequency range while notably suppressing noise in the low-frequency range, in contrast to the conventional method. 
Furthermore, the generation of the noise side is accurate.

\subsection{Addressing the early stopping problem} \label{sec:ex_earlystop}
\begin{figure}[t!]
\begin{center}
\includegraphics[width=6.5cm]{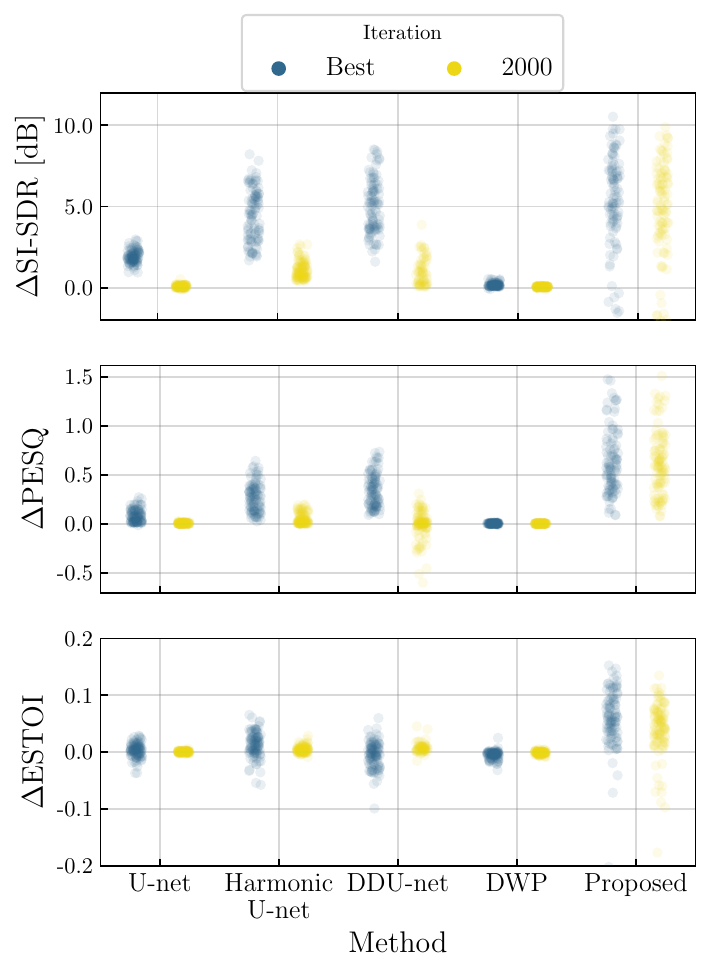}
\caption{
     Scatter plots of the best (i.e., step with the highest SI-SDR) and 2000 steps for each score.
}
\label{fig:ovl_early_stop}
\end{center}   
\end{figure}
Determining the optimal timing for early stopping is a critical concern in conventional DP-based speech enhancement methods, as it directly impacts the performance. 
In principle, the proposed method circumvents this problem by deriving the final output after sufficient training iterations. 
{Figure~\ref{fig:ovl_early_stop}} shows a violin plot of the score against noisy speech with white Gaussian noise for the best and 2000 steps. 
A notable observation here is the significant drop in score for the conventional method when the result at 2000 steps was assumed the processed output. 
Conversely, the proposed method demonstrates similar scores for the best and 2000 steps, indicating a successful resolution of the early stopping problem.

\subsection{Ablation study} \label{sec:ex_ablation}
\begin{table}[tb]
\centering
\label{tb:ablation}
\caption{Results of ablation study}
{\tabcolsep = 1pt
\begin{tabular}{l|ccc}
Method & SI-SDR & PESQ & ESTOI  \\ \hline
Proposed method & $\mathbf{14.83}$ & $\mathbf{2.10}$ & $\mathbf{0.872}$ \\
\quad w/o batch averaging & $14.32$ & $2.04$ & $0.868$ \\
\quad w/o designed DPs & $12.22$ & $1.71$ & $0.800$ \\
\quad w/o Kurtosis loss term & $-3.99$ & $1.07$ & $0.307$ \\
\end{tabular}}
\end{table}
In this section, we assess the effectiveness of each component in the proposed method through an ablation study.
We conducted experiments under identical conditions for three methods:
\begin{itemize}
    \item \textbf{w/o batch averaging}: Excluding batch mean and $\mathcal{L}_2^{(S)}$ from the proposed method.
    \item \textbf{w/o designed DPs}: Utilizing $\beta=2$ softplus function and $\bm{Z}$ from uniform random numbers for both DNNs without designing DPs.
    \item \textbf{w/o kurtosis loss term}: Eliminating $\mathcal{L}_\mathrm{kurt}^{(S)}$ and $\mathcal{L}_\mathrm{kurt}^{(N)}$ from Eq.~(\ref{eq:prop_optim}).
\end{itemize}

{Table~2} summarizes the evaluation scores for each method.
Experimental conditions are the same as those described in section \ref{sec:ex_cond}
Primarily, we observe a rapid deterioration in scores when the kurtosis-based loss term is omitted, underscoring its significant contribution to clean/noise decomposition. Additionally, the exclusion of batch averaging and designed DPs leads to deteriorated scores, 
affirming the integral contribution of these components to the efficacy of the proposed method.

\section{Conclusion}
This paper addresses the limitations of conventional DP-based speech enhancement methods and introduces a new approach that mitigates these issues by leveraging two distinct DNNs along with a spectrogram kurtosis-based loss term. 
Evaluation experiments affirm that the proposed method surpasses existing methods under diverse conditions and adeptly ameliorates the early stopping dilemma prevalent in such frameworks. 
Future enhancements to our method will involve refining the underlying assumptions in the loss term beyond kurtosis. Further improvements can be expected by extending the method from the amplitude domain to the complex and time domains, where additivity holds, provided the complexities of optimization are resolved. 
Moreover, the potential applicability of this method to additional speech processing challenges, such as dereverberation, warrants further investigation.

\section*{Acknowledgements}
This work was partially supported by the Telecommunications Advancement Foundation and the Japan Society for the Promotion of
Science (JSPS) KAKENHI Grant Numbers 24K07513.

\profile{Hien Ohnaka}{
received him B.E degree in 2024. 
At the same time, he graduated from Advanced course of National Institute of Technology, Tokuyama College.
Therefore, he joined the master's course at Nara Institute of Science Technology as a student.
His research interests are in spoken dialogue systems, and speech enhancement.
}
\profile{Ryoichi Miyazaki}{
received the M.E.
and Ph.D. degrees in information science from the
Nara Institute of Science and Technology, Ikoma,
Japan, in 2012 and 2014, respectively.
He is currently a Researcher with the
National Institute of Technology, Tokuyama College, Yamaguchi, Japan.
His research
interests include statistical signal processing and machine learning for speech enhancement.
}


\begin{thebibliography}{99} 
\bibitem{SS} S. Boll, ``Suppression of acoustic noise in speech using spectral subtraction,'' {\em IEEE Trans. on Acoust., Speech, and Signal Process.}, {\bf27}(2), 113--120 (1979).
\bibitem{WF} N. Wiener, ``Extrapolation, interpolation, and smoothing of stationary time series: With engineering applications,'' MIT press (1949).
\bibitem{MMSESTSA} Y. Ephraim and D. Malah, ``Speech enhancement using a minimum-mean square error short-time spectral amplitude estimator,'' {\em IEEE Trans. on Acoust., Speech, and Signal Process.} {\bf32}(6), 1109--1121 (1984).
\bibitem{SE_DNN} Y. Xu, J. Du, L.-R. Dai, and C.-H. Lee, ``A regression approach to speech enhancement based on deep neural networks,'' {\em IEEE/ACM Trans. on Audio, Speech, and Lang. Process.}, {\bf23}(1), 7--19 (2014).
\bibitem{SE_DAE} X. Lu, Y. Tsao, S. Matsuda, and C. Hori, ``Speech enhancement based on deep denoising autoencoder,'' {\em Proc. of Interspeech}, 436--440 (2013).
\bibitem{SE_SEGAN} S. Pascual, A. Bonafonte, and J. Serra ``SEGAN: Speech enhancement generative adversarial network,'' {\em Proc. of Interspeech}, 3642--3646 (2017).
\bibitem{NyTT} T. Fujimura, Y. Koizumi, K. Yatabe, and R. Miyazaki, ``Noisy-target training: A training strategy for DNN-based speech enhancement without clean speech,'' {\em Proc. of EUSIPCO}, 436--440 (2021).
\bibitem{MetricGANU} S.-W. Fu, C. Yu, K.-H. Hung, M. Ravanelli, and Y. Tsao, ``MetricGAN-U: Unsupervised speech enhancement/dereverberation based only on noisy/reverberated speech,'' {\em Proc. of ICASSP}, 7412--7416 (2022).
\bibitem{MixIT} S. Wisdom, E. Tzinis, H. Erdogan, R. Weiss, K. Wilson, and J. Hershey, ``Unsupervised sound separation using mixture invariant training,'' {\em Advances in Neural Info. Process. Systems}, {\bf33}, 3846--3857 (2020).
\bibitem{PULSE} N. Ito and M. Sugiyama, ``Audio signal enhancement with learning from positive and unlabeled data,'' {\em Proc. of ICASSP}, 1--5 (2023).
\bibitem{DAP_harmo} Z. Zhang, Y. Wang, C. Gan, J. Wu, J. B. Tenenbaum,
A. Torralba, and W. T. Freeman, ``Deep audio priors emerge from harmonic convolutional networks,'' {\em Proc. of ICLR}, (2019).
\bibitem{DAP_dense} V. S. Narayanaswamy, J. J. Thiagarajan, and A. Spanias,``On the design of deep priors for unsupervised audio restoration,'' {\em Proc. of Interspeech}, 2167--2171 (2021).
\bibitem{DAP_wave} A. Turetzky, T. Michelson, Y. Adi, and S. Peleg, ``Deep audio waveform prior,'' {\em Proc. of Interspeech}, 2938--2942 (2022).
\bibitem{DAPmus} T. Fujimura and R. Miyazaki, ``Removal of musical noise using deep speech prior,'' {\em Applied Acoust.}, {\bf194}, 108772 (2022).
\bibitem{demucs} A. Defossez, G. Synnaeve, and Y. Adi, ``Real time speech enhancement in the waveform domain,'' {\em Proc. of Interspeech}, 3291--3295 (2020).
\bibitem{mlnfree} R. Miyazaki, H. Saruwatari, T. Inoue, Y. Takahashi, K. Shikano, and K. Kondo, ``Musical-noise-free speech enhancement based on optimized iterative spectral subtraction,'' {\em IEEE Trans. on Audio, Speech, and Lang. Process.}, {\bf20}(7), 2080--2094 (2012).
\bibitem{DIP} D. Ulyanov, V. Lempitsky, and A. Vedaldi, ``Deep image prior,'' {\em International Journal of Computer Vision}, {\bf128}(7), 1867--1888 (2020).
\bibitem{double-dip} Y. Gandelsman, A. Shocher, and M. Irani, ``Double-DIP: Unsupervised image decomposition via coupled deep-image-priors,'' {\em Proc. of CVPR}, 11026--11035 (2019).
\bibitem{JNAS} K. Itou, M. Yamamoto, K. Takeda, T. Takezawa, T. Matsuoka, T. Kobayashi, K. Shikano, and S. Itahashi, ``JNAS: Japanese speech corpus for large vocabulary continuous speech recognition research,'' {\em Journal of the Acoust. Society of Japan (E)}, {\bf20}(3), 199--206 (1999).
\bibitem{DEMAND} J. Thiemann, N. Ito, and E. Vincent, ``DEMAND: A collection of multi-channel recordings of acoustic noise in diverse environments,'' {\em Proc. of Meetings Acoust.}, 1--6 (2013).
\bibitem{U-net} O. Ronneberger, F. Philipp, and B. Tomas, ``U-net: Convolutional networks for biomedical image segmentation,'' {\em Proc. of MICCAI}, 234--241 (2015).
\bibitem{HL} H. Takeuchi, K. Kashino, Y. Ohishi, and H. Saruwatari, ``Harmonic lowering for accelerating harmonic convolution for audio signals,'' {\em Proc. of Interspeech}, 185--189 (2020).
\bibitem{DAP_dense_imple} https://github.com/vivsivaraman/designaudiopriors (accessed 2024-03-21).
\bibitem{DAP_wave_imple} https://github.com/Arnontu/DeepAudioWaveformPrior (accessed 2024-03-21).
\bibitem{InstanceNorm} D. Ulyanov, A. Vedaldi, and V. Lempitsky, ``Instance normalization: The missing ingredient for fast stylization,'' {\em arXiv preprint arXiv:1607.08022}, (2016).
\bibitem{LeakyReLU} A. L. Maas, A. Y. Hannun, A. Y. Ng et al., ``Rectifier nonlinearities improve neural network acoustic models,'' {\em Proc. of ICML}, {\bf30}(1), 3 (2013).
\bibitem{PESQ} J. G. Beerends, A. P. Hekstra, A. W. Rix, and M. P. Hollier, ``Perceptual evaluation of speech quality (PESQ) the new ITU standard for end-to-end speech quality assessment part II: Psychoacoustic model,'' {\em Journal of the Audio Engineer. Society}, {\bf50}(10), 765--778 (2002).
\bibitem{ipcSpec}
Y. Masuyama, K. Yatabe, and Y. Oikawa, ``Low-rankness of complex-valued spectrogram and its application to phase-aware audio processing,'' {\em Proc. of ICASSP}, 855--859 (2019).
\bibitem{DAPsepa} Y. Tian, C. Xu, and D. Li, ``Deep audio prior,'' {\em arXiv preprint arXiv:1912.10292}, (2019).
\bibitem{vad_kurt}
K. Li, M. N. S. Swamy, and M. O. Ahmad, ``An improved voice activity detection using higher order statistics,'' {\em IEEE Trans. on Speech and Audio Process.}, {\bf13}(5), 965--974 (2005).
\bibitem{ica_kurt}
A. Tharwat, ``Independent component analysis: An introduction,'' {\em Applied Computing Informatics}, {\bf17}(2), 222--249 (2021).
\bibitem{ica_moment}
P. Comon, ``Independent component analysis, a new concept?,'' {\em Signal Process.} {\bf36}(3), 287--314 (1994).
\bibitem{SpecKurt} Y. Uemura, Y. Takahashi, H. Saruwatari, K. Shikano, and K. Kondo, ``Automatic optimization scheme of spectral subtraction based on musical noise assessment via higher-order statistics,'' {\em Proc. of IWAENC}, (2008).
\bibitem{DNNkurt} S. Mizoguchi, Y. Saito, S. Takamichi, and H. Saruwatari, ``DNN-Based low-musical-noise single-channel speech enhancement based on higher-order-moments matching,'' {\em IEICE Trans. on Info. and Systems}, {\bf104}(11), 1971--1980 (2021).
\bibitem{sisdr} J. Le Roux, S. Wisdom, H. Erdogan, and J. R. Hershey, ``SDR--half-baked or well done?,'' {\em Proc. of ICASSP}, 626--630 (2019).
\bibitem{estoi} J. Jensen and C. H. Taal, ``An algorithm for predicting the intelligibility of speech masked by modulated noise maskers,'' {\em IEEE/ACM Trans. on Audio, Speech, and Lang. Process.}, {\bf24}(11), 2009--2022 (2016).
\end{thebibliography}
\end{document}